\documentclass[aps,prc,twocolumn,nofootinbib,superscriptaddress]{revtex4-1} 

\usepackage{graphicx,epsfig} %
\usepackage{dcolumn}
\usepackage{mathptmx}
\usepackage{amsmath}
\usepackage{epstopdf}
\usepackage[pdftex]{hyperref}
\usepackage{sidecap}
\sidecaptionvpos{figure}{c}
\usepackage[sort&compress]{natbib}

\usepackage{todonotes}

\usepackage[english]{babel}
\usepackage[utf8x]{inputenc}
\usepackage[T1]{fontenc}

\usepackage[a4paper,top=3cm,bottom=2cm,left=3cm,right=3cm,marginparwidth=1.75cm]{geometry}

\begin{document}
\begin{abstract}
We have made a thorough study of the low-energy behaviour of the $\gamma$-ray strength function 
within the framework of the shell model. We have performed large-scale calculations spanning isotopic and isotonic chains over several mass regions, considering 283 nuclei in total, with the purpose of studying the systematic behavior of the low-energy enhancement (LEE) for $M1$ transitions. There are clear trends in the calculations: From being nearly absent in the lowest mass region, the LEE becomes steeper and more pronounced as the mass number increases, and for a given mass region it further increases towards shell closures. 
Moreover, the LEE is found to be steeper in regions near doubly-magic nuclei where proton particles couple to neutron holes.
These trends enable us to consolidate several previous works on the LEE into a single, consistent concept.
We compare the inferred trends to the available experimental data from the Oslo method, and find support for the systematic behaviour. 
Lastly we have compared the calculations to strength functions compiled from discrete, experimental lifetimes, and find excellent agreement; the discrete data are consistent with an LEE, and indicate that the slope varies as function of mass number.
\end{abstract}

\title{Consolidating the concept of low-energy magnetic dipole decay radiation}

\author{J. E. Midtb{\o}}
\email{j.e.midtbo@fys.uio.no}
\affiliation{Department of Physics, University of Oslo, N-0316 Oslo, Norway}

\author{A. C. Larsen}
\affiliation{Department of Physics, University of Oslo, N-0316 Oslo, Norway}

\author{T. Renstr{\o}m}
\affiliation{Department of Physics, University of Oslo, N-0316 Oslo, Norway}

\author{F. L. Bello Garrote}
\affiliation{Department of Physics, University of Oslo, N-0316 Oslo, Norway}

\author{E. Lima}
\affiliation{Department of Physics, University of Oslo, N-0316 Oslo, Norway}
\maketitle

\section{Introduction}

The atomic nucleus is an extremely complicated many-body quantum system~\cite{al-khalili2005}. Despite intense scrutiny over many decades, many of its facets are still poorly understood. This is especially true when a significant amount of energy is put into the nuclear system, placing it in a highly excited state. Since the number of accessible quantum levels grows approximately exponentially with energy~\cite{bethe1936,ericson1959}, a region of high excitation energy is one where many quantum levels are packed closely together. It is a question of fundamental scientific interest how the quantum-mechanical wave function of such levels is composed, and what degree of correlations exist between the levels \cite{weidenmuller_mitchell}. 

Two basic experimental quantities revealing information on the structure of the nuclear wave functions are excitation-energy levels and their corresponding transition strengths. However, when the excitation energy becomes large, it is experimentally difficult to separate individual levels and transitions, and one instead works with average quantities, such as the energy {\it level density} and $\gamma$-ray {\it strength function}. Our focus in this article is on the strength function, more specifically on the $M1$ component. Evidence for an increasing number of nuclei shows that the $\gamma$-ray strength function exhibits an enhancement towards zero $\gamma$-ray energy (\textit{e.g.}~Refs.~\cite{voinov2004,larsen2018}). This low-energy enhancement (LEE) has been shown to be of dipole order \cite{larsen2013,simon2016,larsen2017,Jones2018}. However, its electromagnetic character is, so far, experimentally undetermined although recent measurements indicate a small bias towards $M1$ transitions \cite{Jones2018}.

The level density and $\gamma$-ray strength function have an important application in calculations of $(n,\gamma)$ capture cross sections 
(\textit{e.g} Ref.~\cite{arnould2007}). Radiative neutron capture is responsible for the synthesis of most elements heavier than iron, mainly through the slow ($s$) and rapid ($r$) neutron-capture processes. The latter process involves neutron-rich nuclei far from stability, close to the neutron drip line. While we are still far from a complete understanding of the $r$ process, which has been singled out as one of the eleven science questions for the 21st century \cite{questions2003}, huge strides were made recently with the discovery of a neutron-star merger event which seemingly produced $r$-process elements \cite{LIGO2017,drout2017,Pian2017}. 
In such a neutron-rich, low-entropy environment, an $(n,\gamma)-(\gamma,n)$ equilibrium cannot be maintained at all times~\cite{arnould2007,eichler2015,mendoza2015}. 
Thus, $(n,\gamma)$ reaction rates become important not only at freeze-out but also for the nucleosynthesis at earlier stages. 
It has been shown that the presence of an LEE in the $\gamma$-ray strength function can impact the $(n,\gamma)$ cross sections by orders of magnitude, especially for neutron-rich nuclei \cite{larsen_goriely}. Hence it is important to obtain an understanding of the prevalence and properties of the LEE.

\section{The history of the low-energy enhancement}\label{sec:LEE_history}
\begin{figure*}[htb]
\centering
\includegraphics[clip,width=\textwidth]{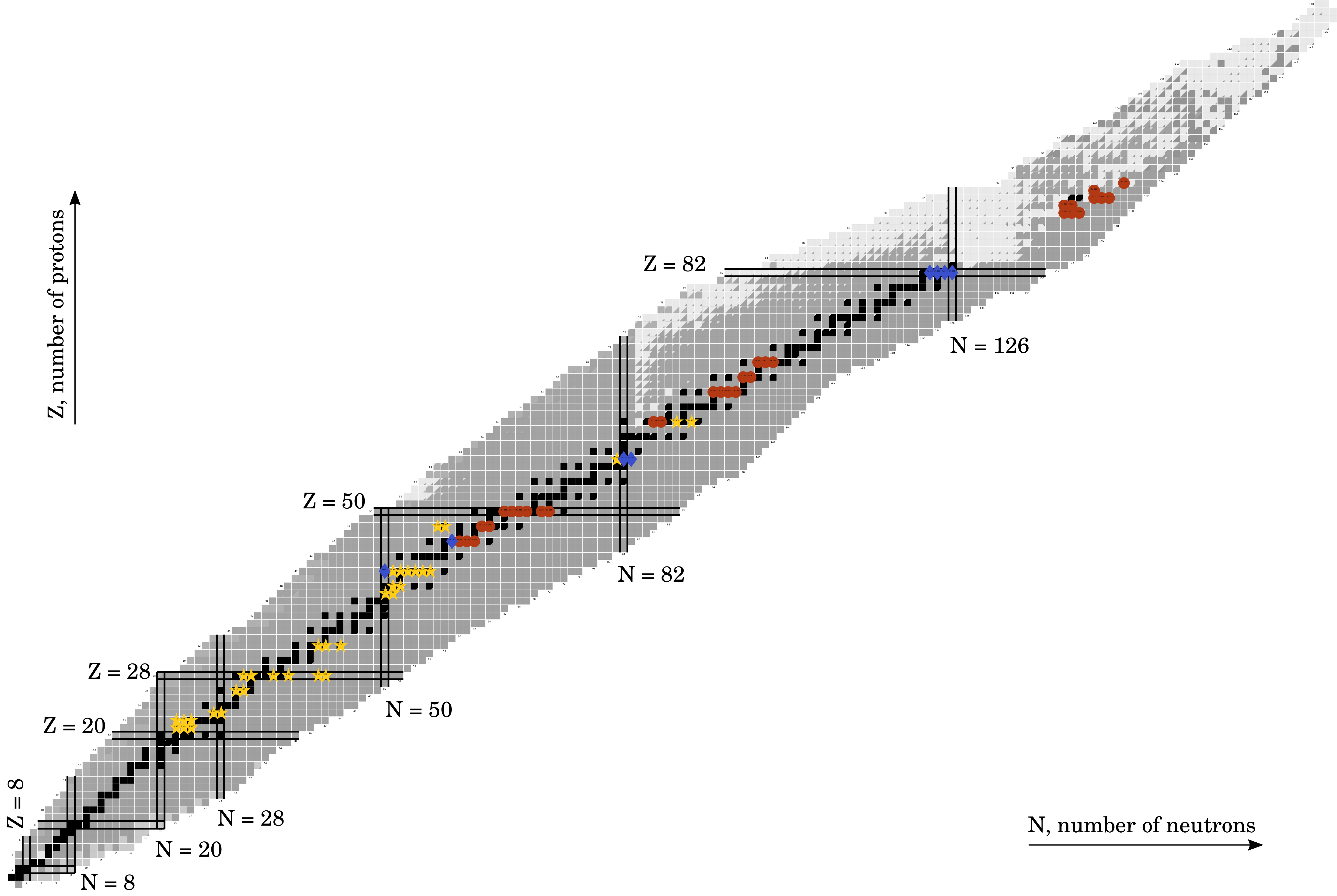}
\caption{\label{fig:oslo_upbends}(Color online) Map detailing where an LEE has been seen using the Oslo method. Yellow stars indicate yes, red circles no. Blue diamonds denote cases where it is difficult to say whether there is an LEE or not. Note that a negative result cannot rule out the presence of an LEE at lower $E_\gamma$ energies than was experimentally accessible (see text for more details). The nuclear chart is made using Ref.~\cite{miernik-chart}, while the experimental data used are from Refs.~\cite{voinov2001,melby2001,siem2002,guttormsen2003,agvaanluvsan2004,voinov2004,guttormsen2005,larsen2006,larsen2007,algin2008,syed2009Pb,syed2009Ti,nyhus2010,toft2010,guttormsen2011,toft2011,burger2012,larsen2013Cd,larsen2014,guttormsen2014,utsunomiya2013,eriksen2014,kheswa2015,spyrou2014,tornyi2014,simon2016,larsen2016,tveten2016,laplace2016,guttormsen2017,renstrom2016,kheswa2017,larsen2017,spyrou2017,larsen2018,crespo2016,crespo2017,larsen2013,renstrom2018Dy,renstrom2018Ni,tveten2018,wiedeking95Mo}.}
\end{figure*}
In Fig.~\ref{fig:oslo_upbends} we have charted the nuclei that have been studied using the Oslo or $\beta$-Oslo methods, and indicated whether the experiment saw a low-energy enhancement or not. 
It must be stressed that experimental limitations make it difficult to extract the very low-$E_\gamma$ strength function using the ($\beta$-)Oslo method. This is mainly due to the uncertainties introduced by unfolding of the Compton-scattering events, which induce large uncertainties the low-$\gamma$ energy spectrum at high excitation energies. Typically, the lower limit on $E_\gamma$ is set at about 1.5 MeV. An exception is $^{151,153}$Sm \cite{simon2016}, where Compton suppression allowed extraction all the way down to $E_\gamma = 700$ keV. In these experiments, they did see a sizable LEE. It could thus be that the LEE is present in some or all of the nuclei marked off with circles and diamonds in the figure. 

Over the last several years, different theoretical interpretations have been put forward to explain the LEE. In fact, the terminology varies, and the phenomenon has been variously referred to as LEE, {\it upbend} \cite{larsen2018}, LEMAR \cite{schwengner2013,schwengner2017} and {\it zero limit} \cite{utsunomiya2018}. If a phenomenon with more than three names can be considered a ``hot topic'', then this clearly qualifies. In the following, we make an attempt to summarize the theoretical work that has been done on explaining the LEE.

Perhaps the first line of demarcation should be drawn between those works explaining the LEE as $M1$ or $E1$ radiation. Litvinova {\it et al.}\ used the thermal-continuum quasiparticle random-phase approximation to demonstrate a low-energy enhancement in the $E1$ strength function \cite{litvinova2013}, introducing a (free) temperature parameter to reproduce the data at low transition energies. On the other hand, a number of authors have explained the LEE as $M1$ radiation by means of shell-model calculations, but with varying interpretations of the underlying mechanism. 

It is difficult to calculate $E1$ strength functions in the shell model, because it requires transitions between wave-function components from different major shells, so-called $1\hbar \omega$ transitions, due to the parity change in the $E1$ selection rule. Inclusion of $1\hbar \omega$ excitations requires a large model space; hence the dimensions of the calculation quickly blow up. It can however be done in some cases, for example by Schwengner \textit{et al.}~\cite{Schwengner2010} and  Sieja \cite{Sieja2017}. 
Still, most shell-model work related to the quasi-continuum strength function to date has been done for $M1$ within $0\hbar \omega$.

The first shell-model study was done by Schwengner \textit{et al.}~\cite{schwengner2013}, who studied Zr and Mo isotopes and compared calculations to strength function data from the Oslo group. They obtained good agreement with the low-energy ($E_\gamma \leq 2$ MeV) $\gamma$-ray strength, and were able to explain almost the complete strength for $E_\gamma < 2$ MeV as being of $M1$ type. They showed that both the distribution of $B(M1)$ values as a function of $E_\gamma$ and the strength function $f_{M1}(E_\gamma)$ can be well fitted by an exponential function, $B_0 \exp (-E_\gamma / T_B)$, with $T_B \sim 0.3-0.5$ MeV and $T_B \sim 0.5$ MeV for $B(M1)$ and $f_{M1}$, respectively.
Further, the mechanism behind the LEE was explained as being due to a recoupling of the spins of high-$j$ protons and neutrons, analogous to the shears-band phenomenon. 

Brown and Larsen \cite{brown2014} investigated the strength function of $^{56,57}$Fe, and were also able to explain it as an $M1$ feature. They further showed that the main contribution to the enhancement is from transition components within orbitals of high $j$, in this case from the $f_{7/2}$ orbital. 

In a subsequent work, Schwengner \textit{et al.}~studied the LEE in a series of Fe isotopes extending into the middle of the neutron shell \cite{schwengner2017}. They found evidence for a {\it bimodality} in the $M1$ strength function, where the total strength is approximately preserved, but the LEE is diminished in the mid-shell isotopes to allow for the emergence of a scissors resonance at $E_\gamma \sim 3$ MeV. Similar to the previous work in Ref.~\cite{schwengner2013}, they stated that the mechanism generating the enhancement is analogous to that of shears bands, \textit{i.e.}~$M1$ transitions generated by a large magnetic dipole moment vector rotating orthogonally to the nuclear spin \cite{frauendorf2001}.

Karampagia {\it et al.}~\cite{Karampagia2017} presented an interesting study using a ``toy model'' where only the $f_{7/2}$ orbital was included, for both protons and neutrons. With this model space they studied $^{49,50}$Cr and $^{48}$V. They again found evidence for a low-energy enhancement, and they showed that its slope is dependent upon the strength of the (in isospin formalism) $T=1$ matrix elements of the nucleon-nucleon interaction.
Like Schwengner \cite{schwengner2013}, they also fitted the $B(M1)$ distribution to an exponential function, but found a much larger $T_B$ of 1.33 MeV, {\it i.e.}~a significantly gentler incline. 

Sieja \cite{Sieja2017} considered the nuclei $^{43,44}$Sc, $^{44,45}$Ti, and obtained both $E1$ and $M1$ strengths by considering a model space comprising three major shells. She found a non-zero low-energy limit of the $E1$ strength function, albeit no enhancement, as the LEE is still explained by the $M1$ component. The $E1$ strength function, although flat, was found to be an order of magnitude weaker than the $M1$ in the low-energy region, thus making no difference to the total strength.

\section{Systematic shell model calculations}
The present work follows the tradition of using the shell model. We employ {\scshape KSHELL} \cite{shimizu2013}, a very efficient $M$-scheme shell model code able to calculate levels and transition strengths within very large model spaces. All the calculations presented here have been made publicly available through Zenodo \cite{zenodo_doi}. As interaction and model space is taken {\scshape jun45} \cite{Honma2009}, which comprises the orbitals $(f_{5/2}pg_{9/2})$ atop a $^{56}$Ni core. The valence space allows up to $22$ protons and neutrons. To facilitate computation, the model space is truncated by turning off proton excitations to the $g_{9/2}$ orbital. We have checked that this does not have an effect on Cu isotopes, but cannot rule out that it could impact nuclei with higher $Z$. Calculations are performed for the entire isotopic chains of Ni, Cu, Zn, Ga, Ge and As that are within the model space, as well as some neutron-rich Se isotopes. For each nucleus, we calculate 100 levels of each parity and each spin between $J=0$ ($J=1/2$) and $J=14$ ($J=29/2$) for even (odd) $A$, respectively. We then calculate $B(M1)$ transition strengths for all allowed transitions and compile the $\gamma$-ray strength function using Eq.~\eqref{eq:gsf}. A bin size of $\Delta E = 0.2$ MeV is used throughout the article unless otherwise stated. For the transition strength calculations in {\scshape jun45}, we use the recommended effective $g_s$ values of $g_{s,\mathrm{eff}} = 0.7 g_{s,\mathrm{free}}$ \cite{Honma2009}.
The dependence of the strength function on $E_x$, $J$ and $\pi$ is removed by averaging. The average includes all calculated states and transitions. 
We observe that the strength function is remarkably similar for different choices of these parameters, except for statistical fluctuations -- hence averaging them out is justified, in accordance with the Brink hypothesis \cite{brink1955}.
\begin{figure}[tb]
\centering
\includegraphics[width=1\columnwidth]{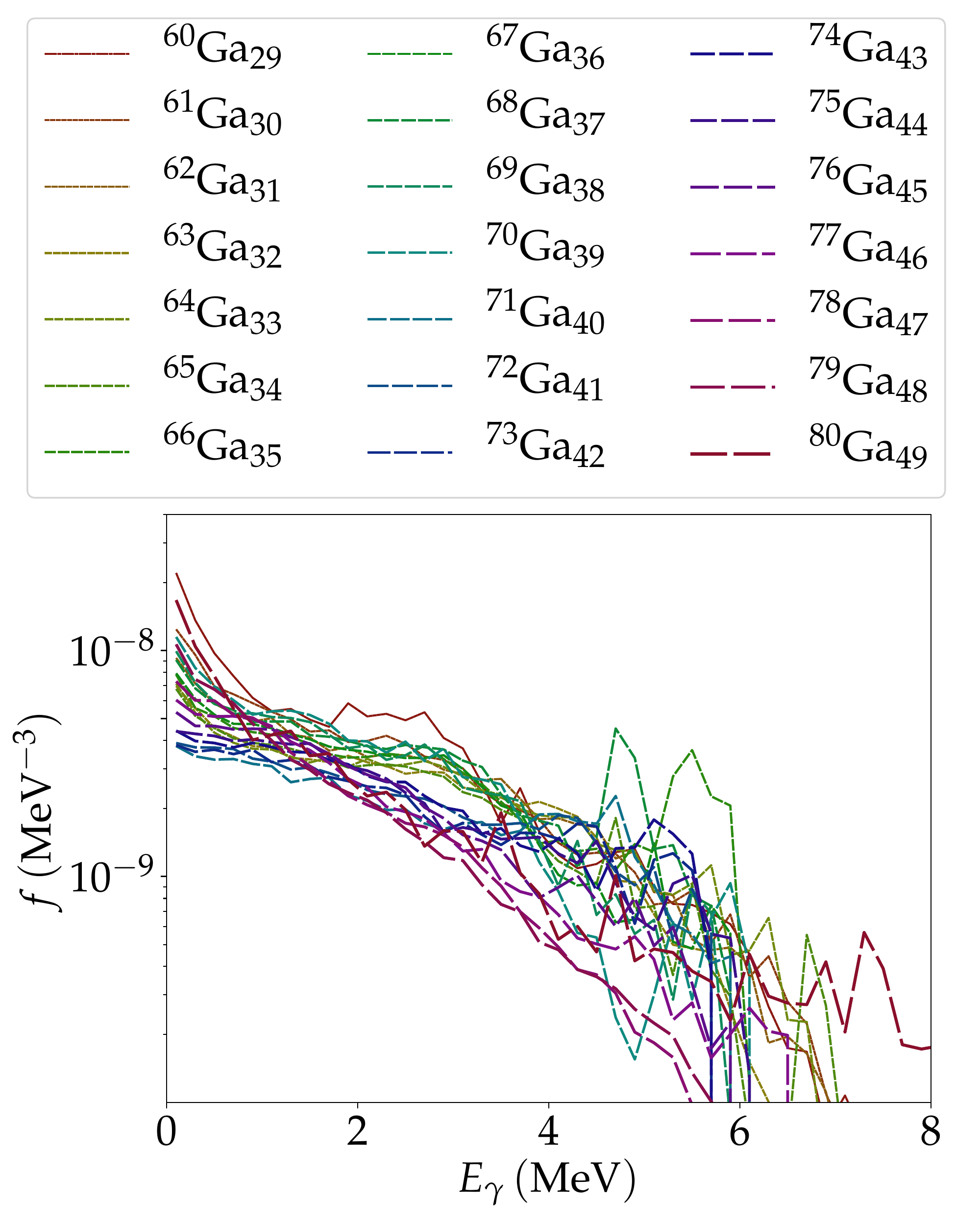}
\caption{\label{fig:Ga_all_gsf}(Color online) Calculated $M1$ $\gamma$-ray strength functions of Ga isotopes using the {\scshape jun45} interaction.}
\end{figure}
As an example, we show the calculated $M1$ strength function of the isotopic chain of Ga isotopes in Fig.~\ref{fig:Ga_all_gsf}.

\begin{figure}[tb]
\includegraphics[width=\columnwidth]{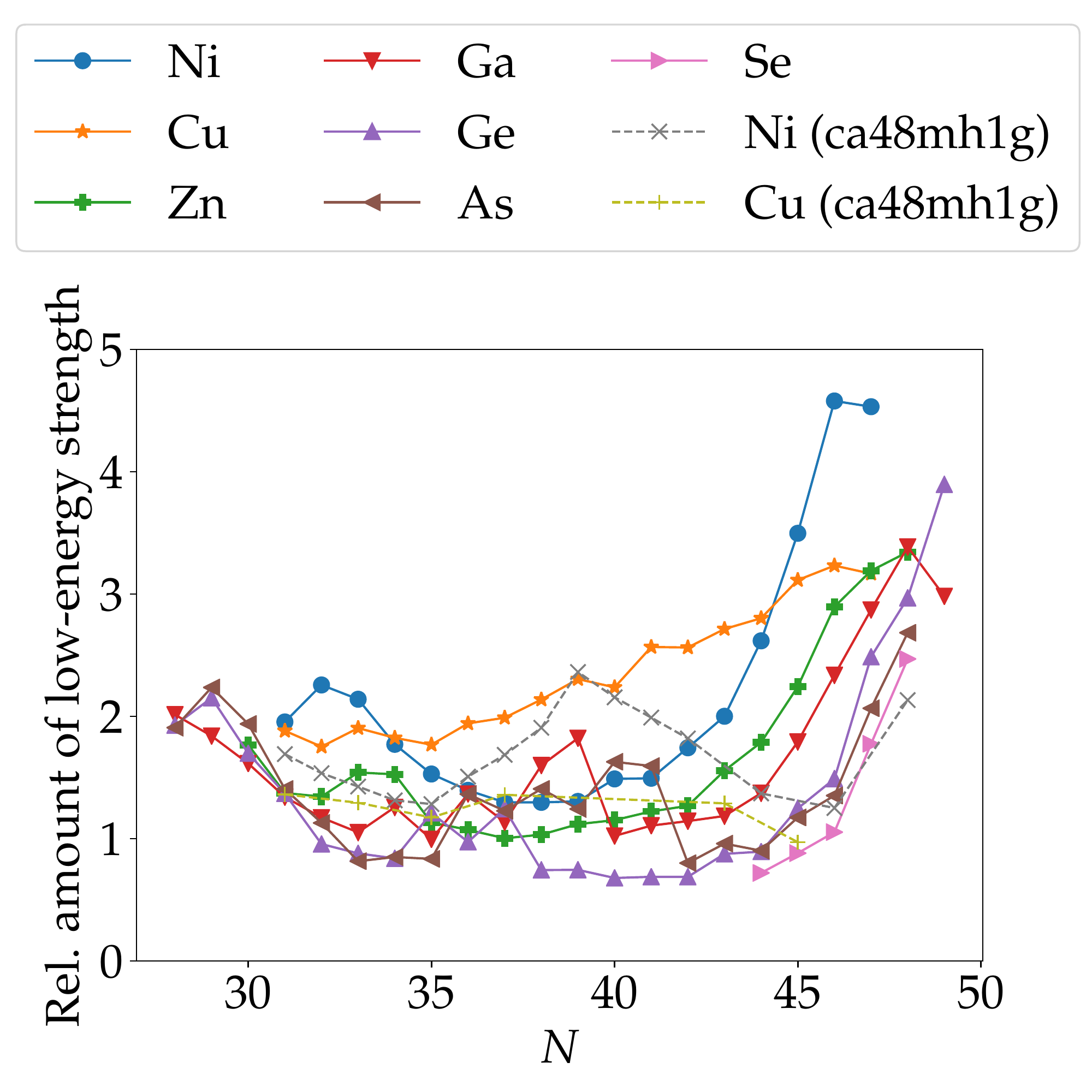}
\caption{\label{fig:correlation_sumrel_jun45}(Color online) The amount of strength between 0 and 2 MeV relative to the strength between 2 and 6 MeV, plotted as function of neutron number for isotopic chains calculated with the {\scshape jun45} and {\scshape ca48mh1g} interactions. See text for details.}
\end{figure}
It is evident from Fig.~\ref{fig:Ga_all_gsf} that the slope changes as function of neutron number. It starts off near $N=28$ being very steep, flattening out towards mid-shell before increasing back again approaching the $N=50$ closure. The same effect is present in the other isotopic chains that we have studied. To see this clearly, we have taken the ratio of the integrated strength in the intervals $E_\gamma \in [0, 2]$ MeV to $E_\gamma \in [2,6]$ MeV, respectively. This is shown in Fig.~\ref{fig:correlation_sumrel_jun45} for all the isotopic chains. The overall trend of increasing low-energy strength towards the shell closures is present for all isotopes. 

\begin{figure}
\includegraphics[width=\columnwidth]{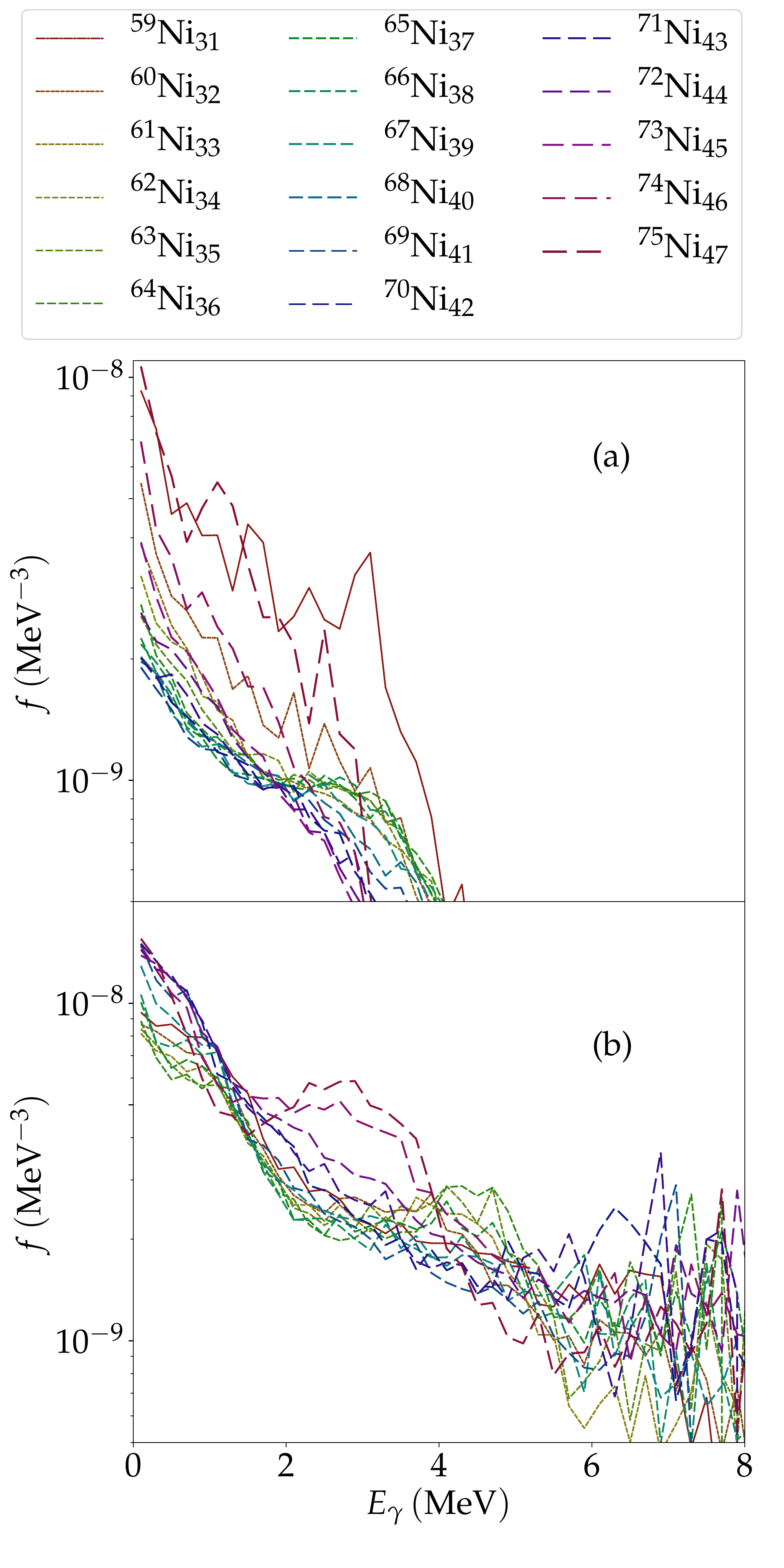}
\caption{\label{fig:Ni_all_gsf}(Color online) $\gamma$-ray strength functions of isotopic chains of Ni calculated with $^{56}$Ni (a) and $^{48}$Ca (b) closed cores, respectively. See text for details.}
\end{figure}

One could worry that some or all of these effects are due to the particulars of the model space, such as the choice of $^{56}$Ni as closed core. In Fig.~\ref{fig:Ni_all_gsf}, we show the chain of Ni isotopes calculated both in the $^{56}$Ni model space and in a different model space, namely using a $^{48}$Ca core with the {\scshape ca48mh1g} interaction \cite{brown_privcomm, larsen2018}, truncated so that two protons can excite from the $f_{7/2}$ orbital. 
Details of the $^{48}$Ca calculations are given in Ref.~\cite{larsen2018}. 
The trend of the strength functions is clearly the same, with more low-energy strength and steeper slope at the shell edges. The 
inclusion of the proton $f_{7/2}$ orbital does however change the strength function, notably by inducing what could be a spin-flip resonance at higher $E_\gamma$ for some of the isotopes. The absolute values are also affected, becoming less variable and generally larger than with the $^{56}$Ni core. 
It is not so surprising that the calculation with only neutrons in the model space gives lower $B(M1)$ values when we consider the structure of the $M1$ operator, $\widehat{M1} \propto g_l \vec{l} + g_s \vec{s}$. Since $g_l^p = 1$, $g_l^n = 0$, the absence of transitions between proton components can lower the strengths.
In Fig.~\ref{fig:correlation_sumrel_jun45}, we have also included the ratio of LEE for the {\scshape ca48mh1g}-calculated Ni isotopes. In this case, the increase at low and high neutron number are complemented by an additional, large bump in the middle, peaking at $^{67}$Ni. The Ni isotopes in the middle of the neutron shell are known to exhibit shape coexistence including spherical components \cite{suchyta2014}. This shape coexistence would involve proton excitations from the $f_{7/2}$ orbital, which means that it should not appear when using the $^{56}$Ni closed core. The {\scshape ca48mh1g} interaction reproduces features attributed to shape coexistence in $^{70}$Ni \cite{larsen2018}. Hence, this mid-shell LEE bump can be interpreted to be consistent with the systematic trends.

Among the {\scshape jun45}-calculated isotopic chains plotted in Fig.~\ref{fig:correlation_sumrel_jun45}, Cu stands out, being linear rather than parabolic as function of $N$. Since Cu has only one proton on top of the $^{56}$Ni core, it is possible that the linear trend is an artifact of the restricted model space. To check this, we again used the {\scshape ca48mh1g} interaction and calculated $^{60,62,64,66,72,74}$Cu, allowing up to two proton excitations from the $f_{7/2}$ as was done for the Ni isotopes. Interestingly, the linearity remains, as shown by the dashed line in Fig.~\ref{fig:correlation_sumrel_jun45}. This seems to indicate that the LEE variation with neutron number is hindered in nuclei with one proton atop magicity. We also note that the same linear trend is present in the fluorine isotopes shown below.

\begin{figure}[tb]
\includegraphics[width=\columnwidth]{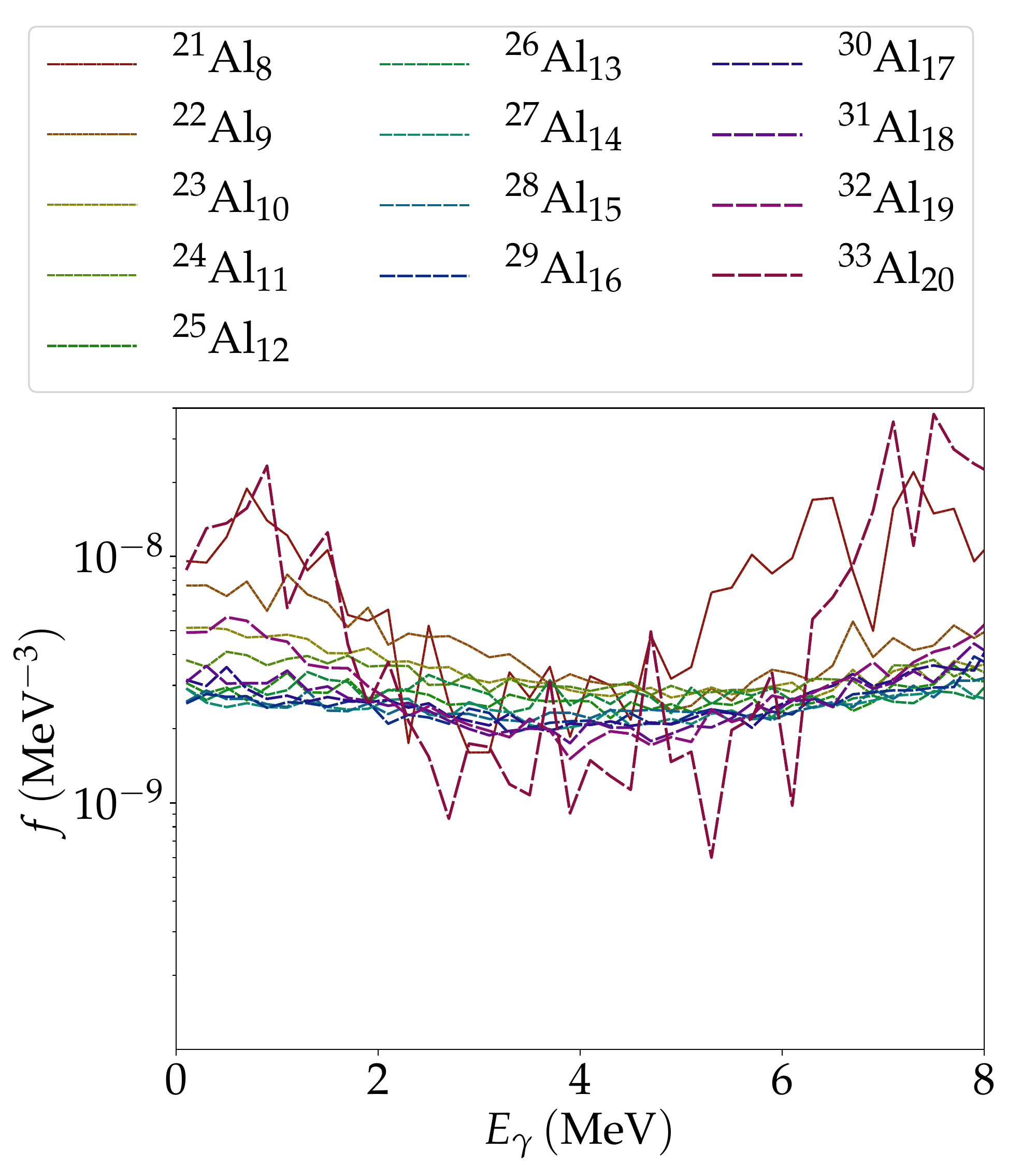}
\caption{\label{fig:Al_all_gsf}(Color online) Calculated $M1$ $\gamma$-ray strength functions of Al isotopes using the {\scshape USDa} interaction.}
\end{figure}

We have made similar calculations as the ones described above in a different mass region, namely the $sd$ shell on top of a $^{16}$O closed core, using the {\scshape USDa} interaction \cite{brown2006}. For this model space, we are able to calculate all isotopes without any truncation. With this interaction, $B(M1)$ strengths are calculated using $g_{s,\mathrm{eff}} = 0.9 g_{s,\mathrm{free}}$ \cite{brown2006}. In Fig.~\ref{fig:Al_all_gsf}, we show the results for the isotopic chain of Al. 
\begin{figure}[tb]
\includegraphics[width=\columnwidth]{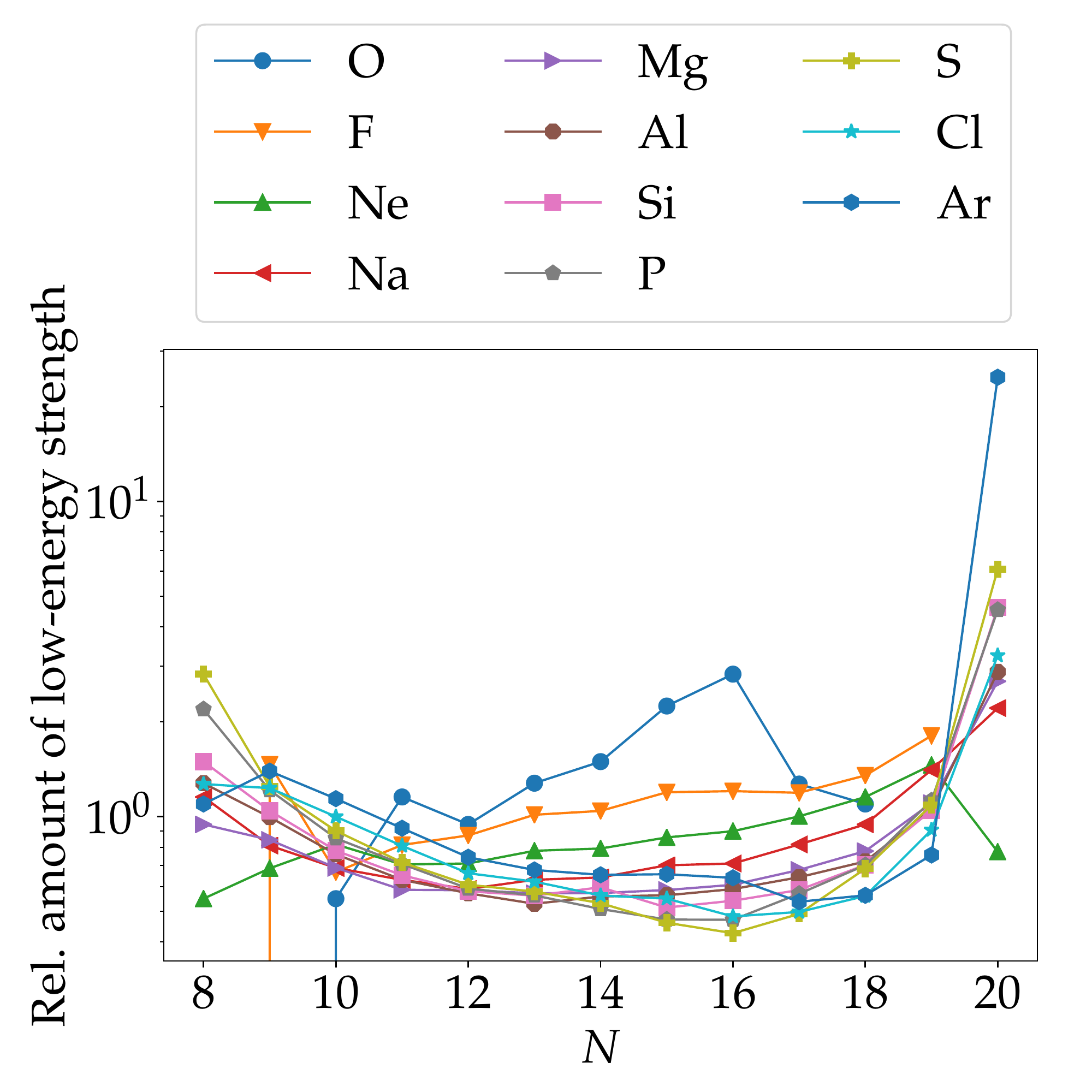}
\caption{\label{fig:correlation_sumrel_usda}(Color online) Correlation between relative sum of low-energy strength and neutron number in the $sd$ region. Note the logarithmic scale.}
\end{figure}
These strength functions are generally much more flat, but reveal the same trend of increase towards magicity. Fig.~\ref{fig:correlation_sumrel_usda} displays the relative amount of low-energy strength for all isotopic chains. There is less change in the LEE as function of $N$ in the middle of the neutron shell compared to the {\scshape jun45} calculations -- but a larger jump at the edges. To make the mid-shell variations more visible, we have used a logarithmic scale.

\begin{figure}[hbt]
\includegraphics[width=\columnwidth]{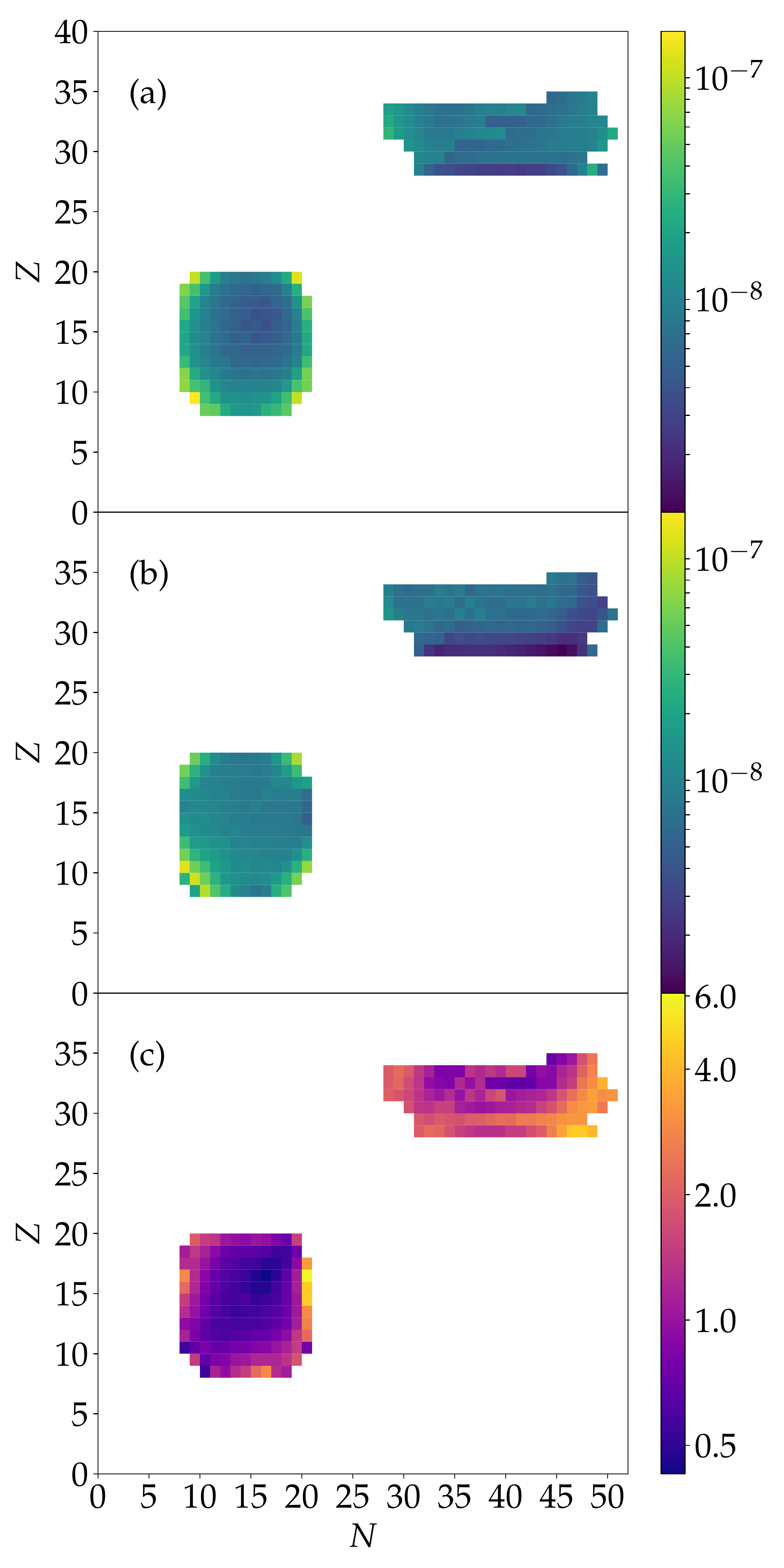}
\caption{\label{fig:calculation_chart}(Color online) Integrated $\gamma$-ray strength from (a) 0 to 2 MeV and (b) 2 to 6 MeV, respectively, and (c) the fraction of the integrated $\gamma$-ray strength from 0 to 2 MeV relative to the 2 to 6 MeV range, \textit{i.e.}~panel (a) divided by panel (b).}
\end{figure}

In Fig.~\ref{fig:calculation_chart}, we have plotted the integrated strength as a nuclear chart. Panel (a) and (b) shows the strength integrated from 0 to 2 MeV and from 2 to 6 MeV, respectively, and (c) shows the ratio between the previous two. This presentation reveals several interesting features. 
First of all, the calculations indicate that the low-energy enhancement is more pronounced near shell closures. Furthermore, the overall steepness of the strength is much higher in the $f_{5/2}pg_{9/2}$ region than the $sd$ region. Lastly, in both model spaces, the southeastern corner 
is enhanced relative to the southwestern one. This is interesting, because it is consistent with the shears band picture advocated in Ref.~\cite{schwengner2017}, as discussed in Section \ref{sec:LEE_history}. We note that the same feature is apparent also in the northern corners of the $sd$ shell, where the north\textit{western} corner has the constructive alignment of proton holes with neutron particles.
Looking at Fig.~\ref{fig:oslo_upbends}, this is consistent with the experimental evidence for nuclei with $A \leq 100$, where an enhancement has been seen in all cases. It is also consistent with the absence of an LEE in the mid-shell regions above $^{132}$Sn and $^{208}$Pb. However, it is seemingly at odds with the data for $^{105-108}$Pd, $^{111,112}$Cd and $^{116-119,121,122}$Sn, where no LEE is seen, despite their proximity to the $Z=50$ shell closure. There could be several explanations for this. It could be that the LEE is very steep, and thus pushed to lower $E_\gamma$ than experimentally accessible. It could also be that the proton shell closure is not a major driving factor for the LEE by itself, or there could be some other mechanism suppressing LEE in this region.

Turning away from the question of relative steepness, it seems, from the present calculations like the $M1$ LEE turns flat rather than disappearing completely, even for the mid-shell $sd$ nuclei. This is important, because it implies that an $M1$ correction term to the $E1$ Lorentzian-like shape typically used in phenomenological models is needed for all nuclei -- but with variable slope. To investigate this point, we have calculated $E1$ strengths for $^{29}$Si. In addition, we have considered $^{44}$Sc, located in the $fp$ shell. The nickel mass region is unfortunately not accessible to $E1$ calculations. We use the {\scshape sdpf-mu} interaction \cite{utsuno2012}, which comprises the $sd$ and $fp$ shells, allowing the cross-shell excitations essential for $E1$ transitions. We have applied a $1\hbar\omega$ truncation, meaning that the single-particle basis configurations are limited to ones where at most one particle is excited across the $sd$-$fp$ shell gap. The Lawson method \cite{bigstick, gloeckner, lawson} with $\beta = 100$ MeV is used to push the spurious centre-of-mass states up to energies outside the considered range. For the $E1$ transition calculations we used effective charges of $e_\mathrm{eff}^p = (1+\chi) e$, $e_\mathrm{eff}^p = \chi e$, with $\chi = -Z/A$ \cite{suhonen}.
\begin{figure}
\includegraphics[width=\columnwidth]{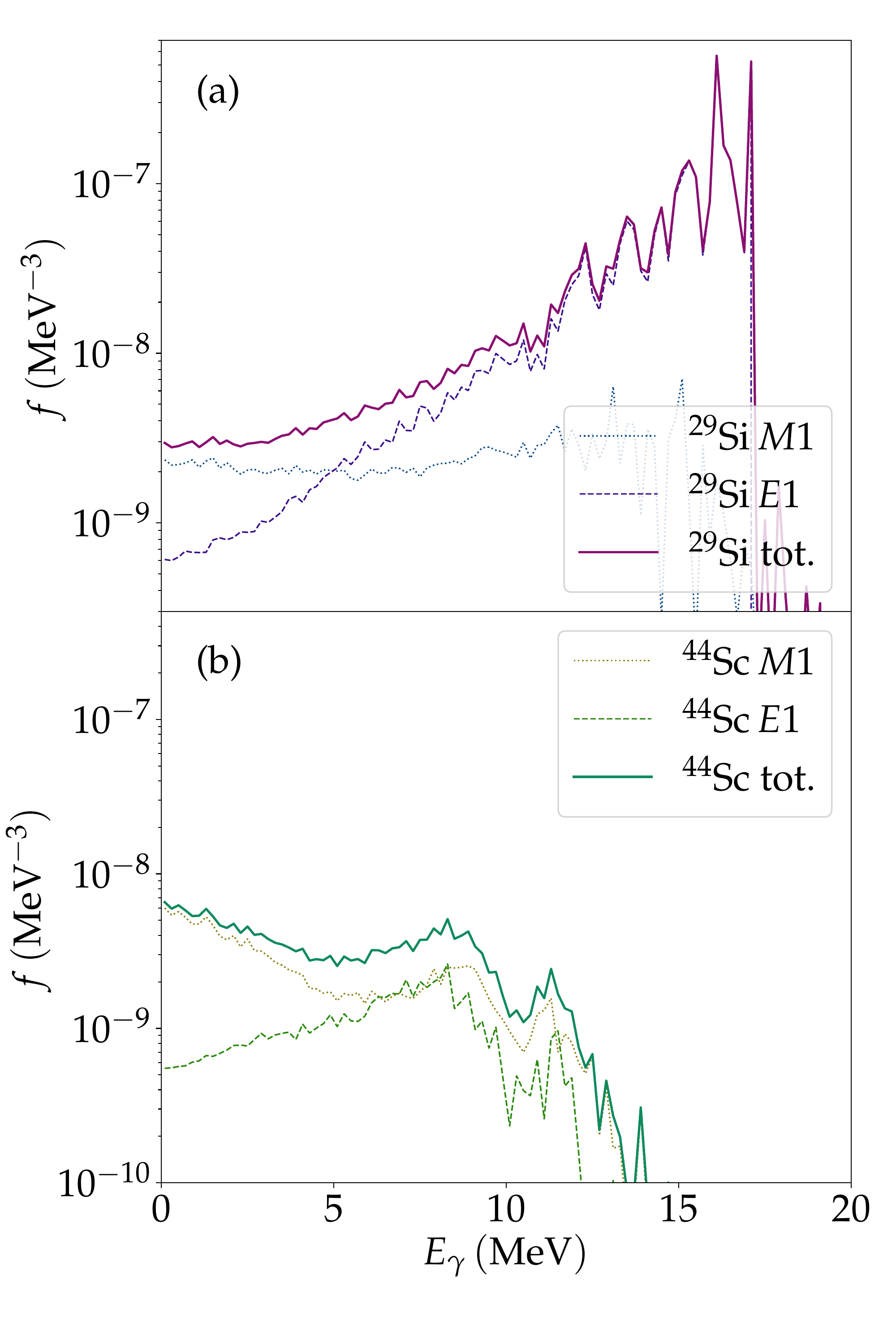}
\caption{\label{fig:M1_and_E1}(Color online) Calculated total dipole strength functions for $^{29}$Si (a) and $^{44}$Sc (b).}
\end{figure}
In both cases, we obtain an $E1$ strength consistent with a Generalized Lorentzian (GLO) tail from the Giant Dipole Resonance (GDR) \cite{kopecky_uhl}. The need for an $M1$ correction is evident in both cases. For $^{29}$Si it only serves to change the slope of the GLO, while for $^{44}$Sc it completely dominates the low-energy part of the strength function, demonstrating an LEE.

Incidentally, we can compare our results with Sieja's calculations for the $E1$ strength in $^{44}$Sc. We find a steeper slope on the low-energy tail of the strength function compared to Fig.~6 in Ref.~\cite{Sieja2017}. This has a large influence on the summed dipole strength function at $E_\gamma \approx 5$ MeV, where we observe a minimum reminiscent of that usually present in the strength function of LEE nuclei. The absolute value of both the $E1$ and $M1$ strength functions are found in the present work to be an order of magnitude lower than in Ref.~\cite{Sieja2017}. This is due to differences in how the strength function is extracted from the $B(E1/M1)$ values (see Appendix A). Both calculations are consistent with the shape of the experimental $\gamma$-ray strength function of $^{44}$Sc from Ref.~\cite{larsen2007}, but Sieja's provide the best match for the absolute value.

\section{Comparisons with discrete experimental data}

Many nuclei are so well studied that we have access to experimental information about levels, lifetimes and branching ratios up to quite high excitation energy. It is interesting to see if this information can be used to compile a strength function, and how it compares to shell model calculations. To this end, we extract experimental information from the RIPL library \cite{ripl-3}. We choose it over other databases due to the ease with which it allows data parsing, despite its lacking transition multipolarity information. 
We thus extract a strength function of {\it presumed} M1 transitions by selecting transitions between levels where $|J_i - J_f| \leq 1$, $\pi_i \pi_f = +1$. 
This does not rule out $E2$ mixing, but based on the power suppression in the multipole expansion, $M1$ is {\it a priori} expected to dominate. As such, this gives an impression of how the low-excitation $M1$ energy strength function behaves. 

For each nucleus considered, we parse the entry in the RIPL library and look for all levels with $E_x \in [0,7]$ MeV that pass the aforementioned requirement and that have a known lifetime and measured $\gamma$-ray branching ratios. From this information we obtain partial decay widths, which we average over $(E_x, E_\gamma, J, \pi)$ bins. The strength function is then obtained by multiplying by the level density at the corresponding $(E_x, J, \pi)$, which we obtain considering {\it all} known levels, not just the ones with known lifetimes. This is important to get the correct absolute value of the strength function (otherwise it would be too low, see Appendix A).
By comparing the level density from the discrete levels to that from shell-model calculations, we verify that the experimental level scheme seems to be complete up to the excitation energies we consider\footnote{If the total level density from RIPL falls below the shell model level density before the ``RIPL used'' density dies off, this would indicate that we are compiling a strength function using too low level density. This does not seem to be the case here.}, as shown in Fig.~\ref{fig:sd_RIPL_NLD}.
Finally, we average over $(E_x,J,\pi)$ to obtain the average strength function depending only on $E_\gamma$. 

\begin{figure}
\includegraphics[width=\columnwidth]{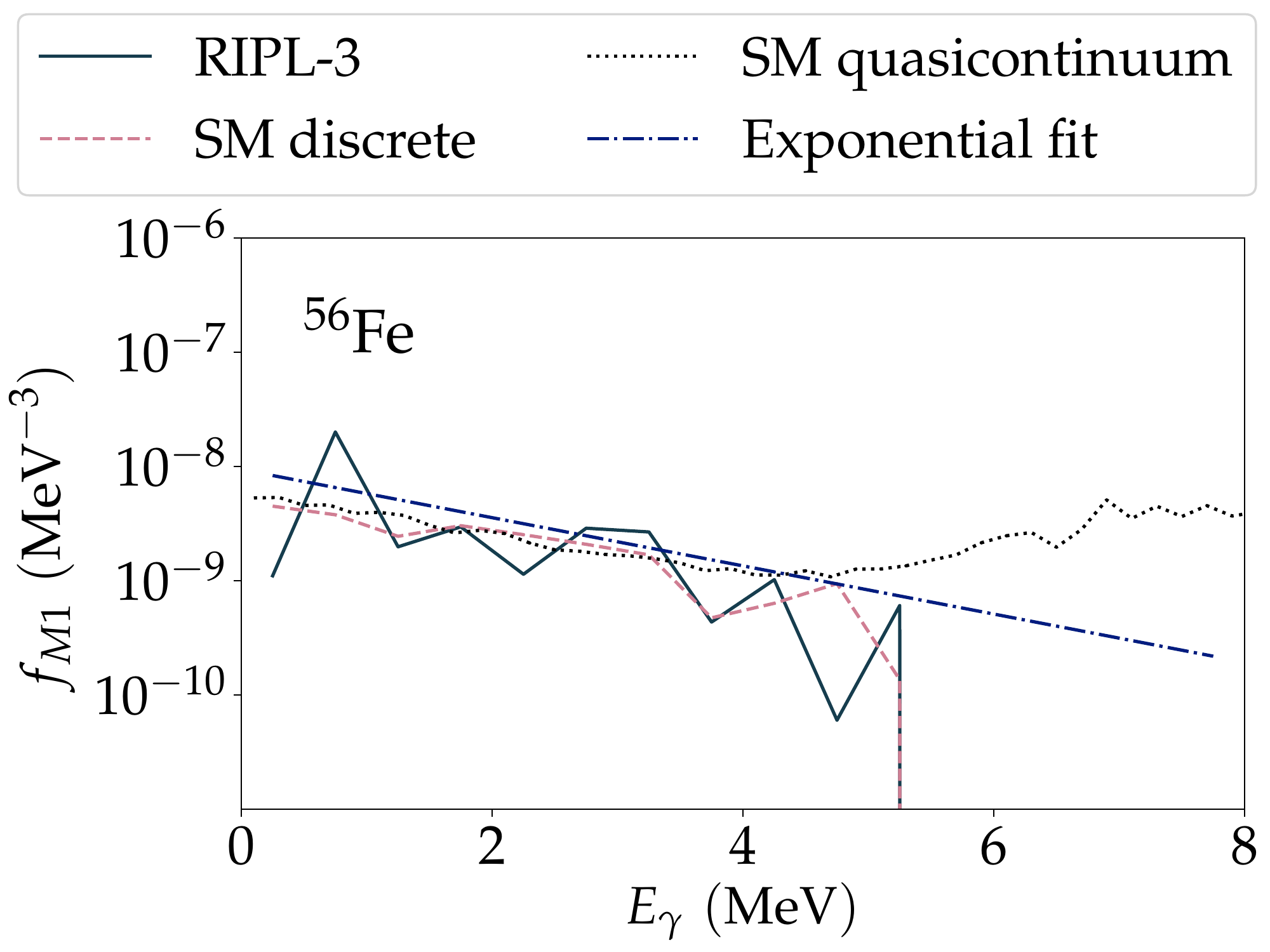}
\caption{\label{fig:RIPL_Fe56}(Color online) Low-energy ``$M1$'' strength function of $^{56}$Fe compiled from discrete experimental data. The bin width is $\Delta E = 0.5$ MeV. See text for details.}
\end{figure}

We demonstrate this for the case of $^{56}$Fe in Fig.~\ref{fig:RIPL_Fe56}. The wealth of available experimental information enables us to construct a strength function based on 90 transitions selected according to the criteria described above. We compare to shell model calculations done using the {\scshape gxpf1a} \cite{gxpf1} interaction, as was used in Ref.~\cite{brown2014}. The agreement between experiment and calculations is excellent, both in terms of slope and absolute value. 
\begin{figure*}[hbt]
\centering
\includegraphics[clip,width=0.99\textwidth]{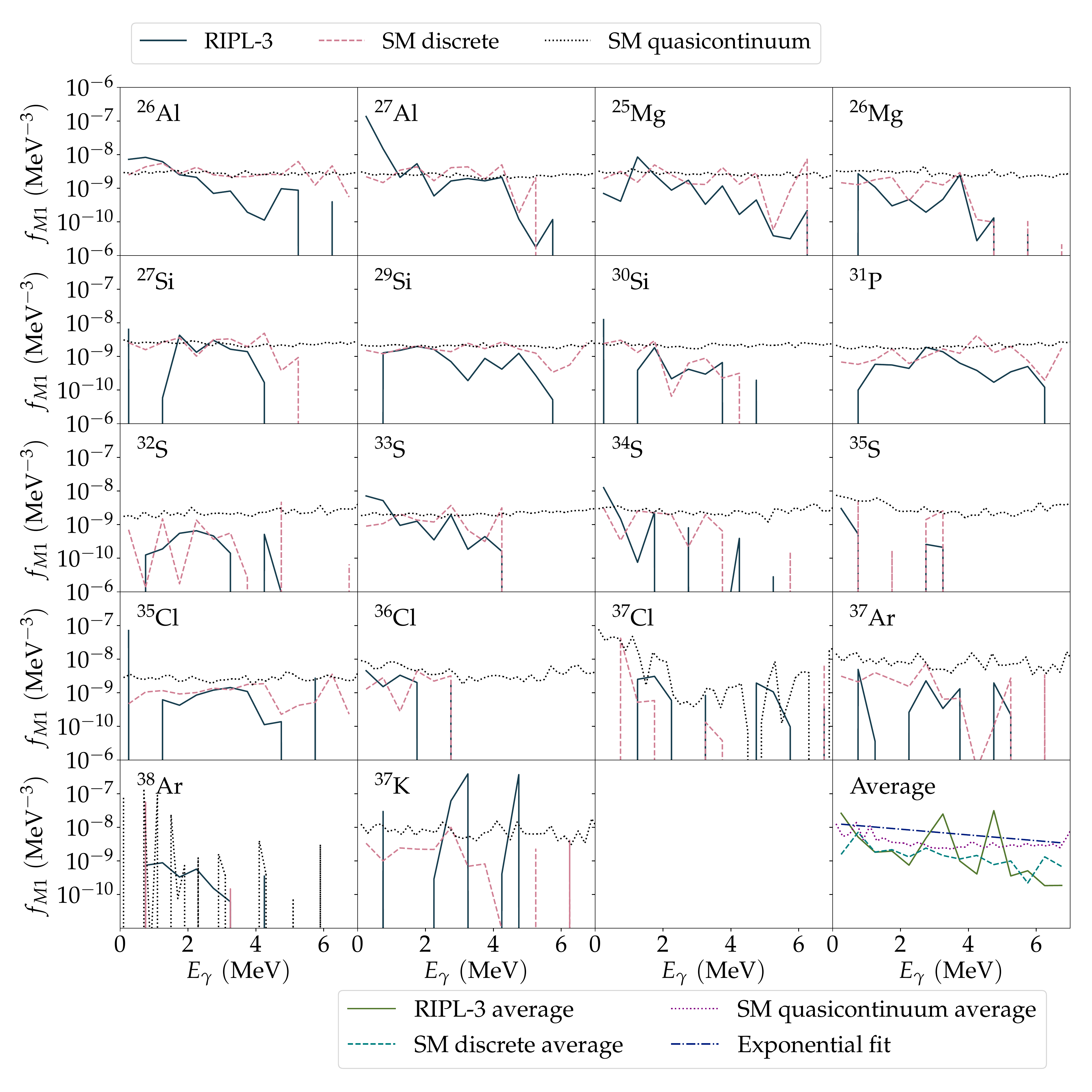}
\caption{\label{fig:SM_vs_RIPL-sd}(Color online) M1 strength function of different $sd$ shell nuclei. The bin width is $\Delta E = 0.5$ MeV.}
\end{figure*}
\begin{figure*}[hbt]
\centering
\includegraphics[clip,width=0.99\textwidth]{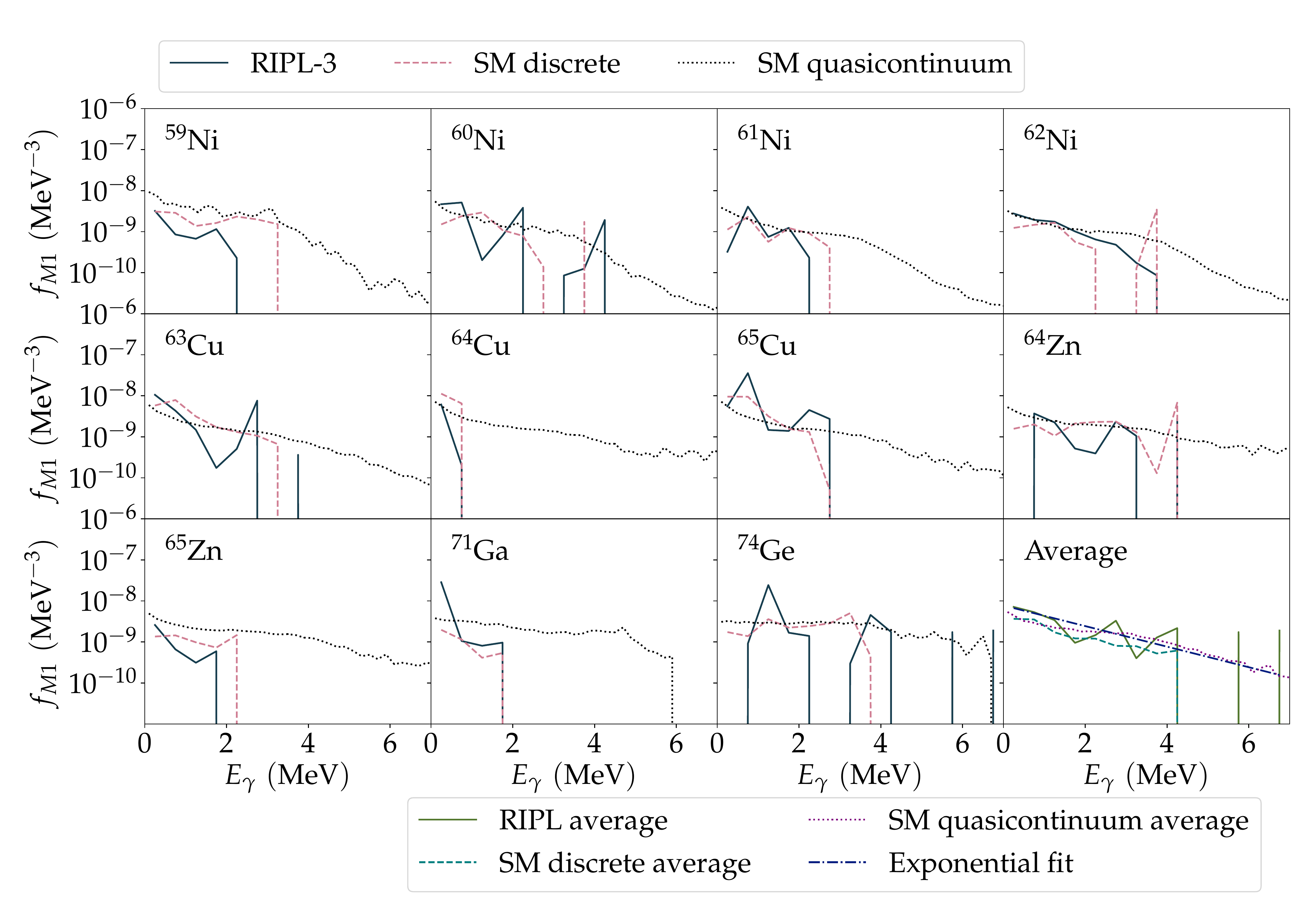}
\caption{\label{fig:SM_vs_RIPL-fpg}(Color online) M1 strength function of different $f_{5/2}pg_{9/2}$-shell nuclei. The bin width is $\Delta E = 0.5$ MeV.}
\end{figure*}
The results for a variety of nuclei in the $sd$ shell
and $f_{5/2}pg_{9/2}$ shell regions are shown in Figs.~\ref{fig:SM_vs_RIPL-sd}
and \ref{fig:SM_vs_RIPL-fpg}, respectively.
\begin{figure}[htb]
\centering
\includegraphics[clip,width=\columnwidth]{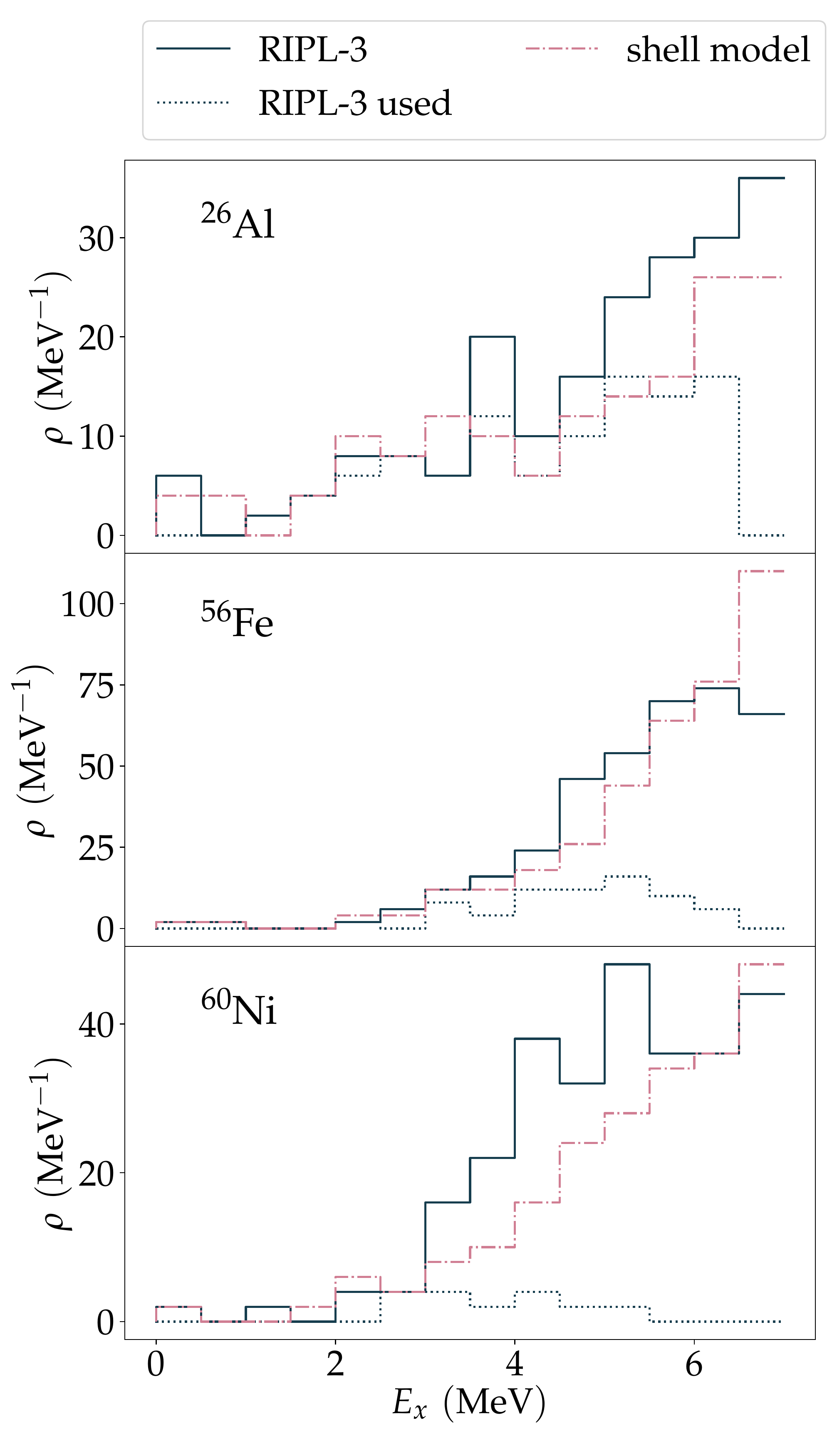}
\caption{\label{fig:sd_RIPL_NLD}(Color online) Level densities from discrete data for different nuclei, compared with shell model calculations. The lines labeled ``RIPL used'' indicate the level density counting only the levels whose lifetimes and branching ratios were used to compile the strength function. The bin width is $\Delta E = 0.5$ MeV.}
\end{figure}
For these regions, we compare to the previously discussed shell model calculations. The dotted line in each strength function panel shows the ``quasi-continuum'' strength function for that nucleus, by which we mean the strength function compiled using all calculated levels, in the same was as was done for the systematics above.
We have also extracted a strength function from the shell model data by selecting discrete transitions similar to the RIPL ones. Specifically, for each RIPL level used in the construction of the strength function, we have taken the lowest-energy 
shell model level with the same spin and parity, and included all transitions from this level in the discrete SM strength function. (We also tried an alternative method selecting the closest-in-energy shell model level, but this gives much poorer results.)

In an attempt to quantify the differences between the mass regions considered, we make a fit to an exponential function $f(E_\gamma) = B \exp(E_\gamma / T)$. To maximize statistics, we fit the average strength function in each of the regions (the green line shown in the last panel of each of the figures). We have also fitted $^{56}$Fe separately. The results for the fit are listed in Table \ref{table:RIPL_fits}. With all the assumptions that go into this fit, we should refrain from drawing strong conclusions, but it is striking that the $sd$ fit displays almost factor 3 gentler slope than $f_{5/2}pg_{9/2}$. This is compatible with the trend from the systematic calculations.
\begin{table}[htb]
\centering
   \begin{tabular}{l l l l l}
        \\
                      & $B \,(10^{-8}$ MeV$^{-3})$ &       $T \,$(MeV)       \\
       \hline
       \hline
       $sd$  				&   1.30   &  5.09   \\
       $f_{5/2}pg_{9/2}$          &   0.77   &  1.73   \\
       $^{56}$Fe  			&   0.94   &  2.07   \\
   \hline
   \end{tabular}
   \caption{Fit parameters for experimental RIPL strength functions. See text for details.}
   \label{table:RIPL_fits}
\end{table}

\section{Summary and outlook}
In this work we have performed large-scale shell model calculations of $M1$ $\gamma$-ray strength functions for many isotopic chains in different major shells, focusing on the low-energy behaviour. We observe systematic trends in the calculations. The slope of the strength functions is generally steeper in the $f_{5/2}pg_{9/2}$ than in the $sd$ shell. This correlates with the availability of high-$j$ orbitals. Furthermore, the slope is steeper near the shell closures and gentler in the mid-shell region for both model spaces. 
This is especially pronounced in the region northwest and southeast of a doubly-magic nucleus, where, in the shears-bands picture, proton and neutron magnetic moments align to generate strong magnetic transitions.

The present findings consolidate several insights from previous studies -- such as the dependence on high-$j$ orbitals, the coupling of protons and neutrons, and the relation to shears bands -- and shows that rather than being separate, incompatible explanations of the low-energy enhancement, they may be complementary pieces of the same puzzle.
Based on this and previous studies, we propose that large low-energy magnetic decay strength is a feature inherent to nuclei when they are excited to high energies.
The slope of the LEE seems to correlate with the availability of high-$j$ orbitals, which also correlates with nuclear mass.
While the slope of the $M1$ strength varies between nuclei and mass regions, it never seems to disappear completely even for the lightest nuclei, but merely turns flat. 
Hence, in phenomenological terms, an $M1$ correction to the strength function at the tail of the $E1$ GDR is probably required for all nuclei, modifying its low-energy shape. 
And indeed, for a large number of them, the low-energy $M1$ strength displays an enhancement. If, as these calculations indicate, the LEE is especially strong for very neutron-rich nuclei, it could significantly impact ($n,\gamma$) reaction rates relevant to the $r$ process.

Whilst there are experimental difficulties preventing definitive exclusions of the LEE with the Oslo method, the data that exist support our present findings. It would be very interesting to study other nuclei in mid-shell regions, and preferably employing experimental techniques enabling the extraction of the strength function to low gamma-ray energy. It is equally interesting to consider nuclei in the ``shears regions'', where we expect the LEE to be most significant. Neutron-rich Xe isotopes are a promising case in this regard, located as they are just northwest of the doubly-magic $^{132}$Sn. An experiment has recently been carried out on $^{133}$Xe at iThemba LABS, and analysis using the Oslo method in inverse kinematics is underway \cite{berg2018}. We eagerly await these experimental results.

\acknowledgments
J.E.M., A.C.L., T.R., and F.L.B.G. gratefully acknowledge financial support through ERC-STG-2014 under grant agreement no. 637686. A.C.L. acknowledges support from the ChETEC COST Action (CA16117), supported by COST (European Cooperation in Science and Technology). Calculations were performed on the Stallo and Fram high-performance computing clusters at the University of Troms{\o}, supported by the Norwegian Research Council. 

All calculations have been made publicly available on Zenodo \cite{zenodo_doi}
\newline

\appendix

\section{Issues with conversion of B(M1) values to strength function}

\begin{figure}[hbt]
\begin{center}
\includegraphics[width=1\columnwidth]{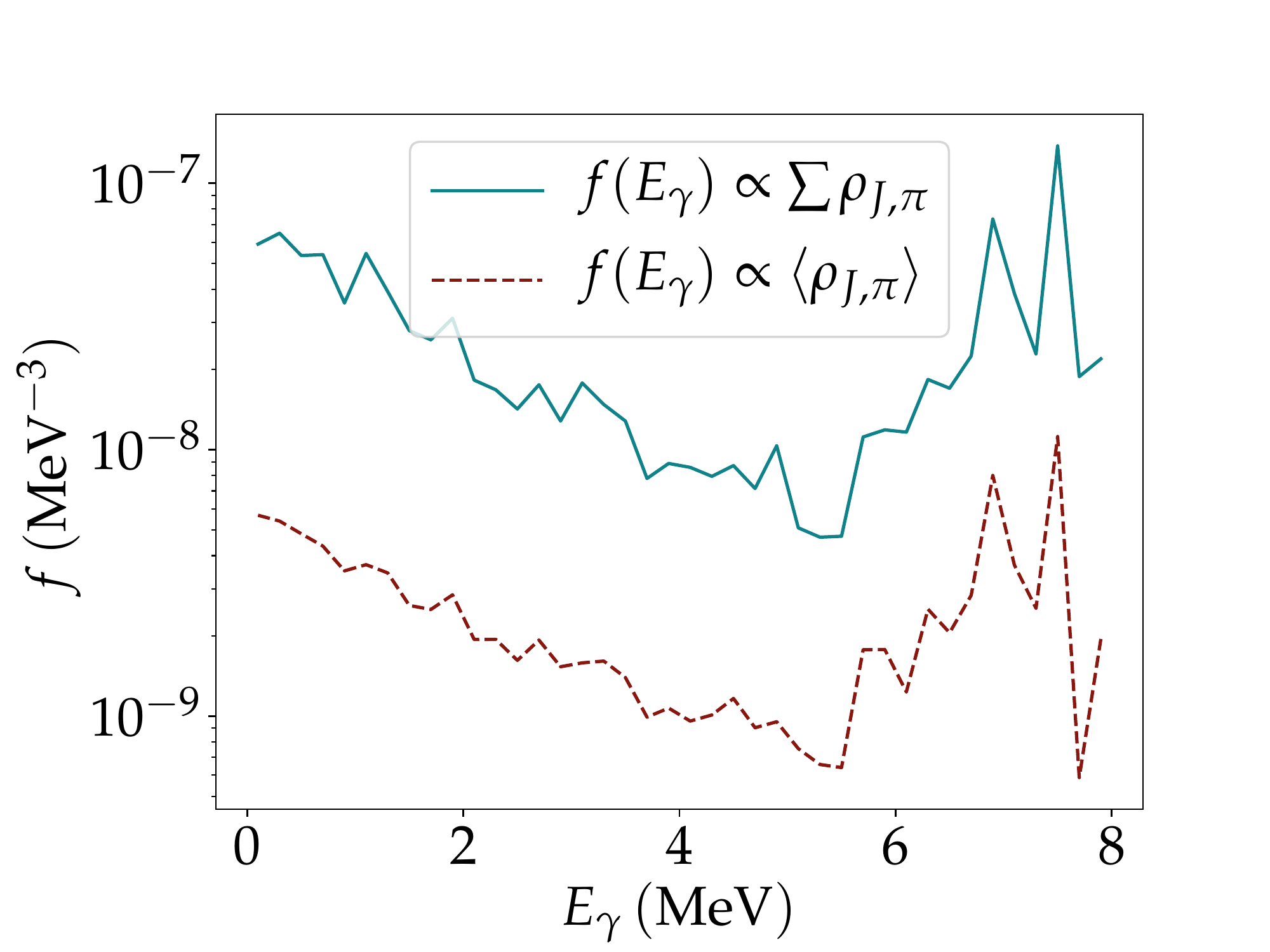}
\caption{\label{fig:Fe56_gsf_method_comparison}(Color online) Comparison of $\gamma$-ray strength functions for $^{56}$Fe from shell model calculations extracted using two different methods. See text for details.}
\end{center}
\end{figure}

We recently became aware of an issue with how shell model calculations are converted to $\gamma$-ray strength functions \cite{gsf_privcomm}. The conventional definition of the strength function, as found in Ref.~\cite{bartholomew1972}, is
\begin{align}
\begin{split}
   \label{eq:gsf} f_{M1}(E_\gamma, &E_i, J_i, \pi_i)\\
   &= \frac{16\pi}{9\hbar^3 c^3} \langle B(M1)\rangle(E_\gamma, E_i, J_i, \pi_i) \rho(E_i, J_i, \pi_i),
\end{split}
\end{align}
where
$\rho(E_i, J_i, \pi_i)$ is the partial level density and $\langle B(M1) \rangle$  is the average transition strength of states at excitation energy $E_i$, spin $J_i$ and parity $\pi_i$. Using that $\mu_N = (e\hbar)/(2m_p c)$, the constant in front works out to 
\begin{align}
	\frac{16\pi}{9\hbar^3 c^3} = 11.58\times 10^{-9}\,\mu_N^{-2} \, \mathrm{MeV}^{-2}.
\end{align}
However, in some works, the \textit{total} level density has been used in place of the partial. Since the total level density is $\rho_\mathrm{tot}(E_x) = \sum_{J,\pi} \rho(E_x, J, \pi)$, this introduces (\textit{i}) 
an artificial overall enhancement of the strength function and (\textit{ii}) an arbitrary scaling depending on how many $J,\pi$ combinations were included in the calculations. In order to demonstrate the difference, we have repeated the calculation of Ref.~\cite{brown2014} and extracted the strength function using both the total and the partial level density. 
In each case we average over $E_x$ and $J$ ($\pi$=+ only). It results in a difference of about a factor 10, as expected since the calculation includes 11 different spins. The effect is demonstrated in Fig.~\ref{fig:Fe56_gsf_method_comparison}. In this work, we keep to the original definition from Ref.~\cite{bartholomew1972}.

\bibliographystyle{alpha}
\bibliography{sample}

\end{document}